\documentclass[a4]{article}
\usepackage{epsfig}

\begin{document}

\title{
The scalar perturbation of\\ the higher-dimensional rotating black holes.
}

\author{
Daisuke Ida,
Yuki Uchida\\
{\it Department of Physics, Tokyo Institute of Technology,}\\
{\it Tokyo 152-8550, Japan}\\
and\\
Yoshiyuki Morisawa\\
{\it Department of Physics, Osaka City University,}\\
{\it Osaka 558-8585, Japan}}

\maketitle

\begin{abstract}

The massless scalar field in the higher-dimensional Kerr black hole (Myers-Perry solution with
a single rotation axis) has been investigated.
It has been shown that the field equation is separable in arbitrary dimensions.
The quasi-normal modes of the scalar field have been searched in five dimensions
using the continued fraction method.
The numerical result shows the evidence for the stability of the scalar perturbation
of the five-dimensional Kerr black holes. The time scale of the resonant oscillation in the rapidly rotating black hole,
in which case the horizon radius becomes small, is characterized by 
(black hole mass)$^{1/2}$(Planck mass)$^{-3/2}$ rather than the light-crossing time of the horizon.

\end{abstract}

\section{Introduction}
The higher dimensional space-times have been more seriously considered 
in recent studies as possible
grounds for the unified theory of elementary particles.
Though there are no experimental evidence that we live in more than four-dimensional
space-time, it is suggested that the existence of the extra dimensions might be proved
by the signature of black hole production in planned high energy experiments.

This argument follows from the brane-world scenario~\cite{brane}.
In such a model, the Standard Model particles are confined to the three-brane,
which means that the Kaluza-Klein modes of the SM particles are not excited,
and only gravitons can propagate in the bulk, hence there are rather mild constraints on the
size of the extra dimensions,
so that the fundamental (higher-dimensional) Planck 
scale can reduce to TeV scale. This is a possible solution to the 
hierarchy problem.
Another type of the brane-world models 
has been proposed by Randall and Sundrum, which utilize the warped metric of the
AdS space.

If the fundamental scale is given by TeV, the black holes will be produced at the 
future colliders such as the CERN Large Hadron Collider (LHC), of which center of mass
energy is 14TeV~\cite{LHC1,LHC2}. 
Though the classical approximation might be marginal
for hypothetical black holes to be produced at LHC,
we shall here discuss the higher-dimensional classical black holes,
which is relevant for future colliders beyond the LHC.
Recent studies show that the higher-dimensional black holes 
might be qualitatively rather different from usual four-dimensional ones.

Significantly, the stationary black hole solutions are shown to be nonunique,
while the uniqueness of static black hole has been established~\cite{unique}.
There are five-dimensional Kerr solution, while Emparan and Reall~\cite{ring} have recently found
the rotating black ring solution with $S^1\times S^2$ horizon, and
there exists a parameter ranges where Kerr black hole and black ring
have the common mass $M$ and the angular momentum $J$.
The dimensionless Kerr parameter $a_*:=a/r_H$, where $a=3J/2M$ is the Kerr parameter and
$r_H$ the horizon radius, are characteristic quantity determining the property of the 
five-dimensional Kerr black hole, which are equivalent to the another dimensionless quantity
$J^2/GM^3$ via
\begin{equation}
{J^2\over GM^3}={32\over 27\pi}{a_*^2\over a_*^2+1}.
\end{equation}
The five-dimensional Kerr black hole can have arbitrary $a_*>0$ for fixed total mass.
Consider a sequence of the solution with the mass being fixed.
Starting with $a_*=0$ (Schwarzschild black hole), spin up the black hole.
At $a_*=2.3$, there emerge two branches of black ring solutions with same $J$, both have
less horizon area than the Kerr black hole. However, for $a_*>2.8$, the horizon area of the
larger black ring exceeds that of the Kerr black hole with
same $J$. Since the horizon area is just the black hole entropy, it means that the
five-dimensional Kerr black hole are entropically unfavourable in microcanonical ensemble.
Though this is a purely thermodynamical argument, one might wonder the stability of the
rapidly rotating black holes, and this is the subject of the present paper.

As a simplest case, we here investigate the massless scalar field in the five-dimensional
Kerr background.
We formulate in arbitrary dimensions for later convenience, 
and then concentrate on the five-dimensional case in actual computation.
The separability in five-dimensional case has been shown in Ref~\cite{frolov}.

\section{Kerr metric in higher dimensions}
The metric of the $(4+n)$-dimensional Kerr metric with only one non-zero angular momentum
is given in Boyer-Lindquist-type coordinates by~\cite{Myers-Perry}
\begin{eqnarray}
g&=&
-{\Delta-a^2\sin^2\vartheta\over\Sigma}dt^2
-{2a(r^2+a^2-\Delta)\over\Sigma}
dtd\varphi \nonumber\\
&&{}+{(r^2+a^2)^2-\Delta a^2 \sin^2\vartheta\over\Sigma}\sin^2\vartheta d\varphi^2
\nonumber\\
&&{}
+{\Sigma\over\Delta}dr^2
+{\Sigma}d\vartheta^2+r^2\cos^2\vartheta d\Omega_n^2,
\label{metric}
\end{eqnarray}
where
\begin{eqnarray}
\Sigma&=&r^2+a^2\cos^2\vartheta,\\
\Delta&=&r^2+a^2-\mu r^{1-n},
\end{eqnarray}
and $d\Omega_n^2$ denotes the standard metric of the unit $n$-sphere.
This metric describes a rotating black hole 
in asymptotically flat, empty space-time with mass and angular momentum
proportional to $\mu$ and $\mu a$, respectively. Hereafter, $\mu,a>0$ are assumed.

The event horizon is located at $r=r_H$, such that $\Delta|_{r=r_H}=0$,
which is homeomorphic to $S^{2+n}$.
When $n=1$, an event horizon exists only when $a<\sqrt{\mu}$, and 
the event horizon shrinks to zero-area in the extreme limit $a\rightarrow\sqrt{\mu}$.
On the other hand, when $n\ge 2$, $\Delta=0$ has exactly one positive root for arbitrary $a>0$.
Thus, there are no extreme Kerr black holes in higher dimensions.

\section{Separation of variables of massless scalar field equations}
Here we consider the scalar perturbation on the background metric (\ref{metric}).
Fortunately, the massless scalar equation turns out to be separable.
We shall just comment that the Hamilton-Jacobi equation for a test particle is also separable, so 
that we have so-called the Carter constant in arbitrary dimensions.

We put $\phi=e^{i\omega t-im\varphi}R(r)S(\vartheta)Y(\Omega)$ into the massless scalar field
equation $\nabla^2\phi=0$, where $Y(\Omega)$ is the hyperspherical harmonics on $n$-sphere,
of which eigenvalues given by $-j(j+n-1)$ ($j=0,1,2,\cdots$).
Then we obtain the separated equations
\begin{eqnarray}
&&{1\over\sin\vartheta\cos^n\vartheta}\left({d\over d \vartheta}
\sin\vartheta\cos^n\vartheta{dS\over d \vartheta}\right)
+\left[\omega^2a^2\cos^2\vartheta
\right.
\nonumber\\
&&{}\left.
-m^2\csc^2\vartheta
-j(j+n-1)\sec^2\vartheta
+A\right]S=0,
\label{ang}
\end{eqnarray}
and
\begin{eqnarray}
&&r^{-n}{d\over d r}\left(r^n\Delta{dR\over d r}\right)
+
\left\{
{\left[\omega(r^2+a^2)-ma\right]^2\over\Delta}
\right.
\nonumber\\
&&{}\left.
-{j(j+n-1)a^2\over r^2}
-\lambda
\right\}R=0,
\label{rad}
\end{eqnarray}
where $\lambda:=A-2m\omega+\omega^2 a^2$.

We consider the eigenvalue problem of Eqs.~(\ref{ang}) and (\ref{rad})
subject to the quasi-normal boundary condition.
This boundary condition is given by
\begin{eqnarray}
&&R\sim (r-r_H)^{i\sigma},~~~(r\rightarrow r_H)\nonumber\\
&&R\sim r^{-(n+2)/2}e^{-i\omega r},~~~(r\rightarrow +\infty)
\end{eqnarray}
where 
\begin{equation}
\sigma:={(r_H^2+a^2)\omega-ma\over (n-1)(r_H^2+a^2)+2r_H^2}
\end{equation}
has been determined by the asymptotic behavior of the Eq.~(\ref{rad}).
In other words, the waves are purely ingoing at the horizon and purely outgoing at the
infinity.

\subsection{Possible Unstable Modes}
Here we show along the line of 
Detweiler and Ipser~\cite{Detweiler-Ipser}
that the possible unstable mode is restricted to some region of the
complex $\omega$-plane.
We assume that Eqs.~(\ref{ang}) and (\ref{rad}) have an unstable eigenmode with frequency $\omega$:
\begin{equation}
\omega=\omega_R+i\omega_I, \quad \omega_I<0,
\end{equation}
where $\omega_R$ and $\omega_I$ are the real and imaginary part of $\omega$, respectively.
Multiplying scalar field equation $\nabla^2\phi=0$ by 
$r^n\sin\vartheta\cos^n\vartheta\Sigma\phi^{*}$, 
and integrating over the region outside the event horizon,
 we obtain
\begin{eqnarray}
&&\int\int\left\{-2\omega_R
\left[{\Sigma(r^2+a^2)+\mu a^2 r^{1-n}\sin^2\vartheta \over \Delta}\right]
+m{2a\mu r^{1-n} \over \Delta}\right\}|\phi|^2r^n\sin\vartheta\cos^n\vartheta dr d\vartheta=0,
\label{eq:DetweilerIm}
\end{eqnarray}
and

\begin{eqnarray}
&&\int\int\left\{
\left[j(j+n-1)\left(\sec^2\vartheta+{a^2\over r^2}\right)
+(\omega_I^2-\omega_R^2)
\left[{\Sigma(r^2+a^2)+\mu a^2 r^{1-n}\sin^2\vartheta \over \Delta}\right]
\right.\right.
\nonumber\\
&&\left.\left.+m\omega_R{2a\mu r^{1-n}\over\Delta}
+m^2\left(\csc^2\vartheta-{a^2\over\Delta}\right)\right]|\phi|^2
+\Delta\left|{\partial\phi\over\partial r}\right|^2+\left|{\partial\phi\over\partial\vartheta}\right|^2
\right\}
r^n\sin\vartheta\cos^n\vartheta dr d\vartheta=0.
\label{eq:DetweilerRe}
\end{eqnarray}
The Equation~(\ref{eq:DetweilerIm}) implies that
 $m\omega_R>0$ holds when $m\ne0$, and $\omega_R=0$ when $m=0$.
Since $\Delta\ge0$ is monotonically increasing function for $r>r_H$, we have
\begin{eqnarray}
r^{n-1}\left[\Sigma(r^2+a^2)+\mu a^2 r^{1-n}\sin^2\vartheta\right]&&\nonumber\\
=r^{n-1}\left[(r^2+a^2)^2-a^2\Delta\sin^2\vartheta \right]&\ge& 
r^{n-1}\left[(r^2+a^2)^2-a^2\Delta\right]\nonumber\\
&=&r^{n+3}+r^{n+1}a^2+\mu a^2\nonumber\\
&>&r_H^{n+3}+r_H^{n+1}a^2+\mu a^2
\end{eqnarray}
Then, from Eq.~(\ref{eq:DetweilerIm}), the real part of $\omega$ turns out to be
bounded:
\begin{equation}
|\omega_R|<{|m|a\mu\over r_H^{n+3}+a^2r_H^{n+1}+\mu a^2}
={|m|a \over \mu r_H^{1-n}} :=|m|\omega_H.
\end{equation}
On the other hand, using a similar inequality
\begin{eqnarray}
r^{n-1}[(r^2+a^2)^2-a^2\Delta\sin^2\vartheta]
&\ge& r^{n+3}+r^{n+1}a^2+\mu a^2\nonumber\\
&\ge& r_H^{n+3}+r_H^{n+1}a^2+\mu a^2,
\end{eqnarray}
Eq.~(\ref{eq:DetweilerRe}) leads to the inequality
\begin{eqnarray}
&&\omega_I^2\le\omega_R^2+{m^2 a\over r_H^2}\omega_H,\quad (n\ge1),\\
&&\omega_I^2\le(|\omega_R|-|m|\omega_H)^2,\quad (n=1).
\end{eqnarray}
In particular, it turns out that there no unstable mode
in
the axisymmetric ($m=0$) and the Schwarzschild ($a=0$) cases.
In the non-axisymmetric case, when unstable modes are not ruled out,
we rely on the numerical calculation.

\section{Numerical Computation}
Here we employ the  continued fraction method ~\cite{Leaver,Leaver2,Gautschi}~for this problem, 
which can determine the resonant frequency $\omega$ and the separation constant $A$
with very high accuracy.
We assume the following series expansion for $S$
\begin{eqnarray}
S&=&(\sin\vartheta)^{|m|}(\cos\vartheta)^{j}\sum_{k=0}^\infty a_k(\cos^2\vartheta)^k,
\end{eqnarray}
which automatically satisfies the regular boundary conditions at $\vartheta=0,\pi/2$
whenever converges. Substituting this into Eq.~(\ref{ang}), we obtain the 
three-term recurrence relations
\begin{eqnarray}
&&\alpha_0a_1+\beta_0a_0=0,\\
&&\alpha_ka_{k+1}+\beta_ka_k+\gamma_ka_{k-1}=0,~~~(k=1,2,\cdots)
\label{angrec}
\end{eqnarray}
where
\begin{eqnarray}
\alpha_k&=&-8(k+1)(2j+n+2k+1),\\
\beta_k&=&2\left[(j+|m|+2k)(j+n+|m|+2k+1)-A\right],\\
\gamma_k&=&-\omega_*^2a_*^2.
\end{eqnarray}
Here we have defined the dimensionless quantity $\omega_*:=\omega r_H$ and 
$a_*:=a/r_h$, since the behavior of the system depends only on $a_*$.
When $a_*=0$, the eigenvalue $A$ is explicitly determined from the requirement that
the series expansion ends within finite terms, since otherwise divergent. Thus we have
\begin{eqnarray}
A&=&(2\ell+j+|m|)(2\ell+j+|m|+n+1)+O(\omega_*a_*),\nonumber\\
&&(\ell=0,1,2,\cdots)
\end{eqnarray}
and the 0th-order eigenfunctions are given in terms of the Jacobi polynomials:
\begin{eqnarray}
P_{\ell jm}&=&(\sin\vartheta)^{|m|}(\cos\vartheta)^{j}\nonumber\\
&&\times F\left(-\ell,\ell+j+|m|+{n+1 \over 2},j+{n+1 \over 2};\cos^2\vartheta\right).\nonumber\\
\end{eqnarray}

In a similar way, we expand the radial function $R$ into the form
\begin{eqnarray}
R&=&e^{-i\omega_*r}
\left({r-r_H\over r_H}\right)^{i\sigma}
\left({r+r_H\over r_H}\right)^{-(n+2)/2-i\sigma}
\nonumber\\
&&{}\times \sum_k^\infty b_k\left({r-r_H\over r+r_H}\right)^k,
\label{radepd}
\end{eqnarray}
which automatically satisfies the quasi-normal boundary condition,
whenever convergent.

When $n=1$, substituting Eq.~(\ref{radepd}) into Eq.~(\ref{rad}), we obtain
the five-term recurrence relations
\begin{eqnarray}
\tilde\alpha_k b_{k+1}
+\tilde\beta_k b_{k}
+\tilde\gamma_k b_{k-1}
+\tilde\delta_k b_{k-2}
+\tilde\epsilon_k b_{k-3}=0,
\\
(k=0,1,\cdots)\nonumber
\end{eqnarray}
where
$b_{-1}=b_{-2}=\cdots=0$ and coefficients are given by
\begin{eqnarray}
\tilde\alpha_k 
&=&4(k+1)(k+1+2i\sigma),\\
\tilde\beta_k
&=&-6-4k-4(k+1)^2a_*^2-4\lambda+32\omega_*\sigma
\nonumber\\
&&{}+i\left[2ma_*-(16k+10+2a_*^2)\omega_*\right]
,\\
\tilde\gamma_k
&=&
-8k^2+4k-11+8j^2a_*^2-8\lambda+64\omega_*\sigma
\nonumber\\
&&{}-4i(4k-1)(2\omega_*+\sigma)
,\\
\tilde\delta_k
&=&
-8+4k-4j^2a_*^2-4\lambda+32\omega_*\sigma
\nonumber\\
&&{}-4i\left[
4(k-1)\omega_*-\sigma
\right]
,\\
\tilde\epsilon_k
&=&
+(2k-3)(2k-3+4i\sigma)
.
\end{eqnarray}
For $n\ge2$, the recurrence relations have more terms.
In what follows, we restrict ourselves on the $n=1$ case
as a simple case.
The five-term recurrence relations reduce to the four-term ones 
\begin{equation}
\tilde\alpha'_kb_{k+1}+\tilde\beta'_kb_k+\tilde\gamma'_kb_{k-1}
+\tilde\delta'_kb_{k-2}=0,~~~(k=0,1,\cdots)
\end{equation}
via the Gaussian
elimination:
\begin{eqnarray}
\tilde\alpha_1'&=&\tilde\alpha_1,~~~
\tilde\beta_1'=\tilde\beta_1,~~~
\tilde\gamma_1'=\tilde\gamma_1,\\
\tilde\alpha_2'&=&\tilde\alpha_2,~~~
\tilde\beta_2'=\tilde\beta_2,~~~
\tilde\gamma_2'=\tilde\gamma_2,~~~
\tilde\delta_2'=\tilde\delta_2,\\
\tilde\alpha_k'&=&\tilde\alpha_k,~~~
\tilde\beta_k'=\tilde\beta_k-\tilde\alpha'_{k-1}\tilde\epsilon_k/\tilde\delta'_{k-1},
\nonumber\\
\tilde\gamma_k'&=&\tilde\gamma_k-\tilde\beta'_{k-1}\tilde\epsilon_k/\tilde\delta'_{k-1},~~~
\tilde\delta_k'=\tilde\gamma_k-\tilde\beta'_{k-1}\tilde\epsilon_k/\tilde\delta'_{k-1},
\nonumber\\
&&(k=3,4,\cdots)
\end{eqnarray}
The similar procedure
\begin{eqnarray}
\tilde\alpha_1''&=&\tilde\alpha_1',~~~
\tilde\beta_1''=\tilde\beta_1',~~~
\tilde\gamma_1''=\tilde\gamma_1',\\
\tilde\alpha_k''&=&\tilde\alpha_k',~~~
\tilde\beta_k''=\tilde\beta_k'-\tilde\alpha''_{k-1}\tilde\delta'_k/\tilde\gamma''_{k-1},
\nonumber\\
\tilde\gamma_k''&=&\tilde\gamma_k'-\tilde\beta''_{k-1}\tilde\delta'_k/\tilde\gamma''_{k-1},
~~~(k=2,3,\cdots)
\end{eqnarray}
leads to the three-term relations
\begin{equation}
\tilde\alpha''_kb_{k+1}+\tilde\beta''_kb_k+\tilde\gamma''_kb_{k-1}=0.
~~~(k=0,1,\cdots)
\label{radrec}
\end{equation}
Given the three-term recurrence relations 
the formal expressions of the ratio of the successive coefficients are
obtained in two ways:
\begin{eqnarray}
{a_k\over a_{k-1}}
&=&
-{\beta_{k-1}\over\alpha_{k-1}}
-{1\over\alpha_{k-1}}
{-\gamma_{k-1}\alpha_{k-2}\over\beta_{k-2}+}
\cdots
{-\gamma_{2}\alpha_{1}\over\beta_{1}+}
{-\gamma_1\alpha_0\over\beta_0}
\nonumber\\
&=&-{\gamma_k\over\beta_k+}
{-\alpha_k\gamma_{k+1}\over\beta_{k-1}+}
{-\alpha_{k+1}\gamma_{k+2}\over\beta_{k-2}+}
\cdots,
\label{angconfra}
\end{eqnarray}
where $(1/x+)(y/z)$ abbreviates $1/[x+(y/z)]$.
In a similar manner, we obtain
\begin{eqnarray}
{b_{k'}\over b_{k'-1}}&=&-{\tilde\beta''_{k'-1}\over\tilde\alpha''_{k'-1}}
-{1\over\tilde\alpha''_{k'-1}}
{-\tilde\gamma''_{k'-1}\tilde\alpha''_{k'-2}\over\tilde\beta''_{k'-2}+}
\cdots
{-\tilde\gamma''_2\tilde\alpha''_1\over\tilde\beta''_1+}
{-\tilde\gamma''_1\tilde\alpha''_0\over\tilde\beta''_0}
\nonumber\\
&=&-{\tilde\gamma''_{k'}\over\tilde\beta''_{k'}+}
{-\tilde\alpha''_{k'}\tilde\gamma''_{k'+1}\over\tilde\beta''_{k'-1}+}
{-\tilde\alpha''_{k'+1}\tilde\gamma''_{k'+2}\over\tilde\beta''_{k'-2}+}
\cdots.
\label{radconfra}
\end{eqnarray}
For fixed $k$ and $k'$, we have simultaneous equations
(\ref{angconfra}) and (\ref{radconfra}) for two unknown
complex numbers $\omega_*$ and $A$.
We perform the numerical evaluation of the solutions using a standard nonlinear root 
search algorithm provided by the MINPACK subroutine package.
The  equations for different sets of ($k$,$k'$) are used as a numerical check.

\begin{figure}
\centerline{\epsfbox{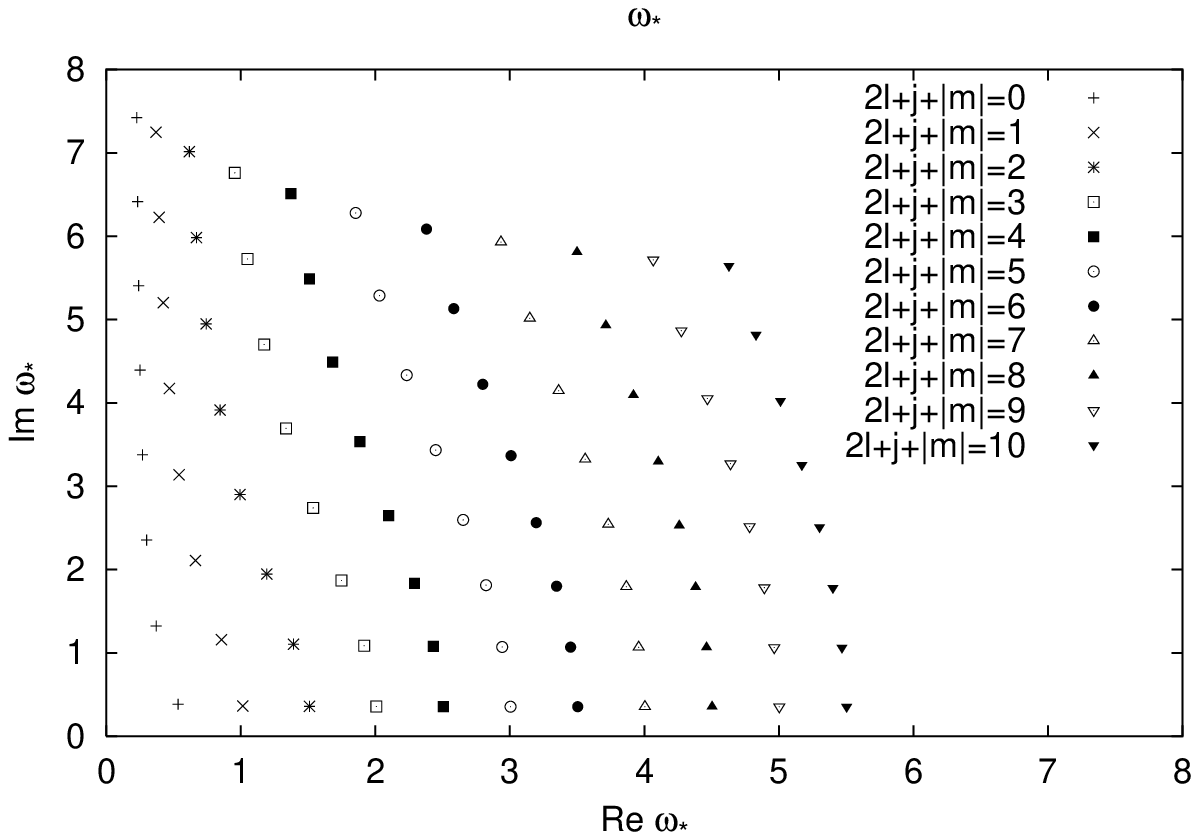}}
\caption{For non-rotating black hole,
first eight quasi-normal frequencies for $2\ell+j+|m|=0,\cdots,10$
are plotted in the complex $\omega_*$ plane.}
\label{fig:Schw}
\end{figure}

In the case of non-rotating black hole, the first eight quasi-normal modes with several values of angular momentum obtained in our computation are plotted in Fig.~\ref{fig:Schw}.

\begin{figure}
\centerline{\epsfbox{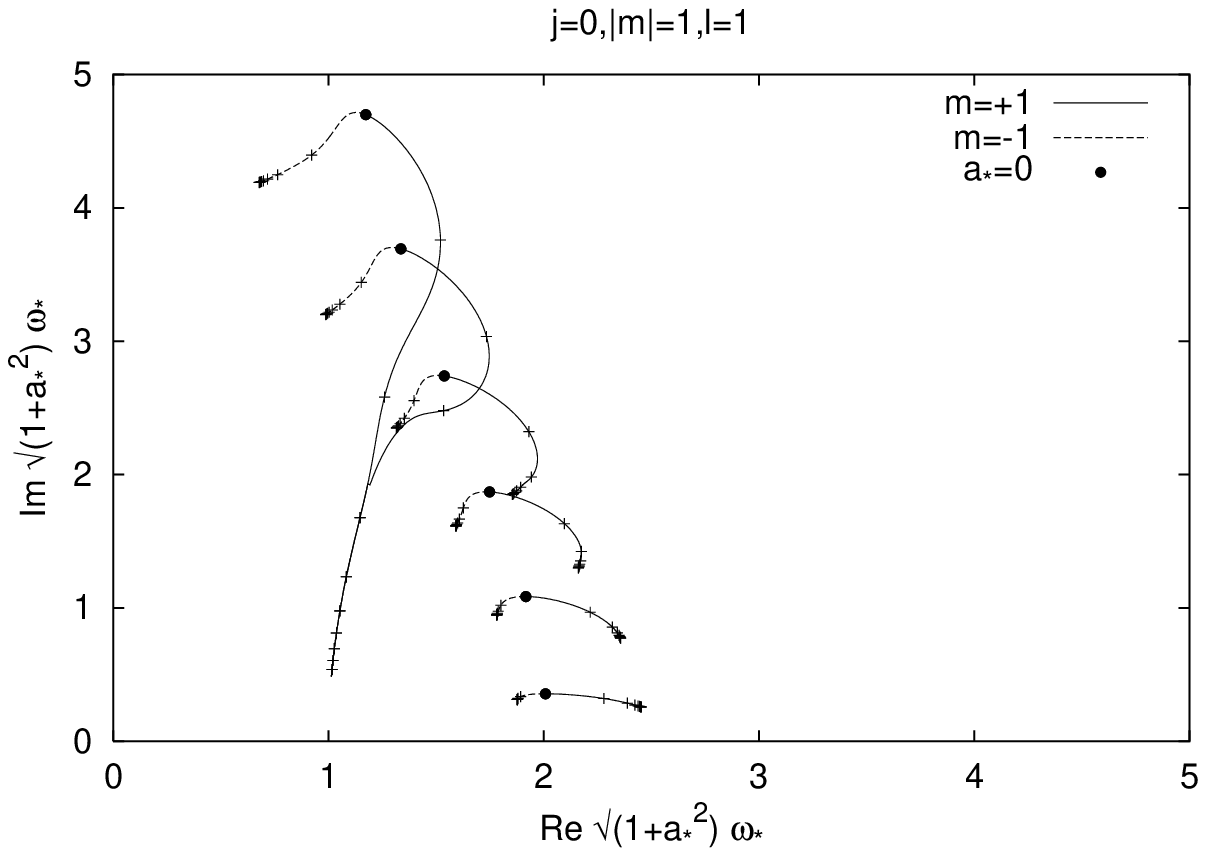}}
\caption{The first six quasi-normal modes of Kerr black hole for $j=0$, $m=\pm1$, $\ell=1$ are plotted in the $\sqrt{1+a_*^2}\omega_*$ plane.
The bullets are the quasi-normal modes of Schwarzschild black hole ($a_*=0$).
The right part starting from the Schwarzschild mode is the branch of the $m=+1$ modes.
The left part is the branch of the $m=-1$ modes.
The crosses are the modes with integer values of $a_*$.
}
\label{fig:011}
\end{figure}

\begin{figure}
\centerline{\epsfbox{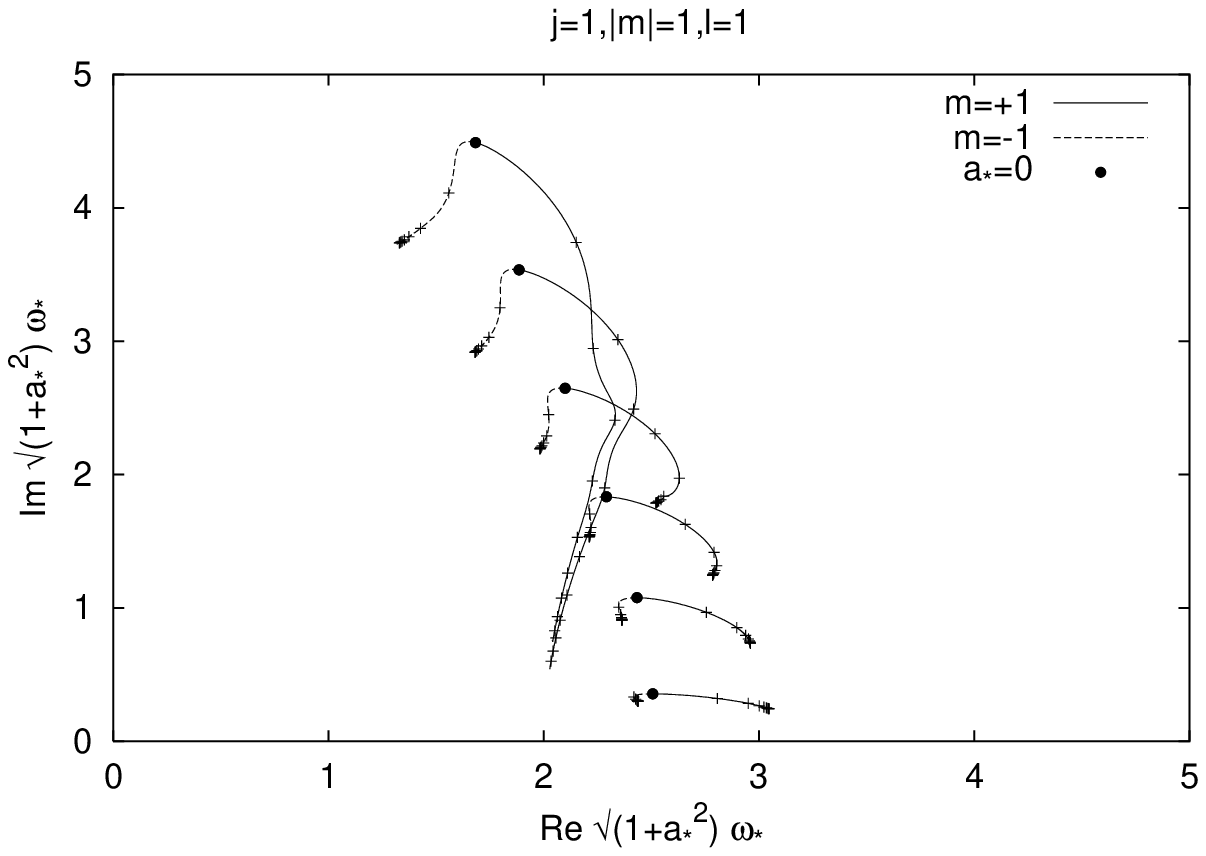}}
\caption{The first six quasi-normal modes of Kerr black hole for $j=1$, $m=\pm1$, $\ell=1$ are plotted in the $\sqrt{1+a_*^2}\omega_*$ plane.
The bullets are the quasi-normal modes of Schwarzschild black hole ($a_*=0$).
The right part starting from the Schwarzschild mode is the branch of the $m=+1$ modes.
The left part is the branch of the $m=-1$ modes.
The crosses are the modes with integer values of $a_*$.
}
\label{fig:111}
\end{figure}

\begin{figure}
\centerline{\epsfbox{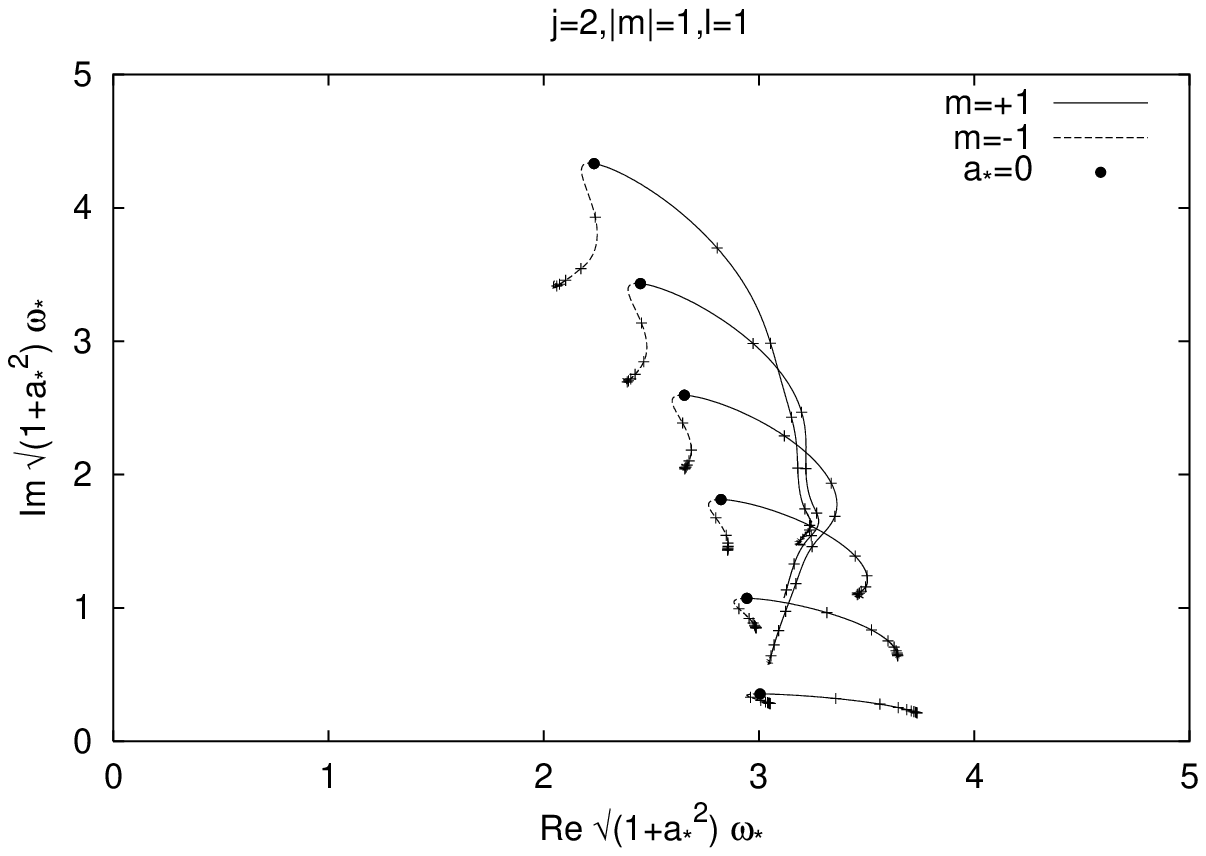}}
\caption{The first six quasi-normal modes of Kerr black hole for $j=2$, $m=\pm1$, $\ell=1$ are plotted in the $\sqrt{1+a_*^2}\omega_*$ plane.
The bullets are the quasi-normal modes of Schwarzschild black hole ($a_*=0$).
The right part starting from the Schwarzschild mode is the branch of the $m=+1$ modes.
The left part is the branch of the $m=-1$ modes.
The crosses are the modes with integer values of $a_*$.
}
\label{fig:211}
\end{figure}

\begin{figure}
\centerline{\epsfbox{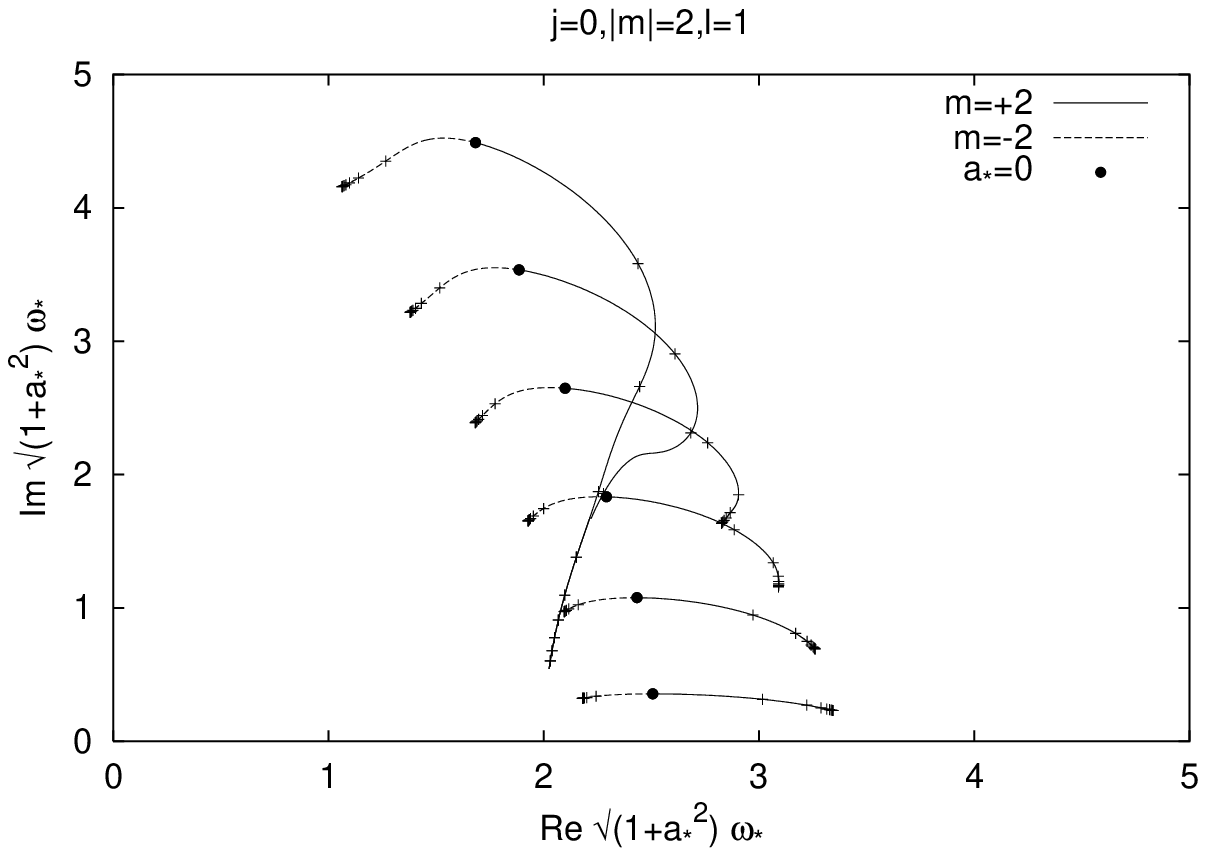}}
\caption{The first six quasi-normal modes of Kerr black hole for $j=0$, $m=\pm2$, $\ell=1$ are plotted in the $\sqrt{1+a_*^2}\omega_*$ plane.
The bullets are the quasi-normal modes of Schwarzschild black hole ($a_*=0$).
The right part starting from the Schwarzschild mode is the branch of the $m=+2$ modes.
The left part is the branch of the $m=-2$ modes.
The crosses are the modes with integer values of $a_*$.
}
\label{fig:021}
\end{figure}

\begin{figure}
\centerline{\epsfbox{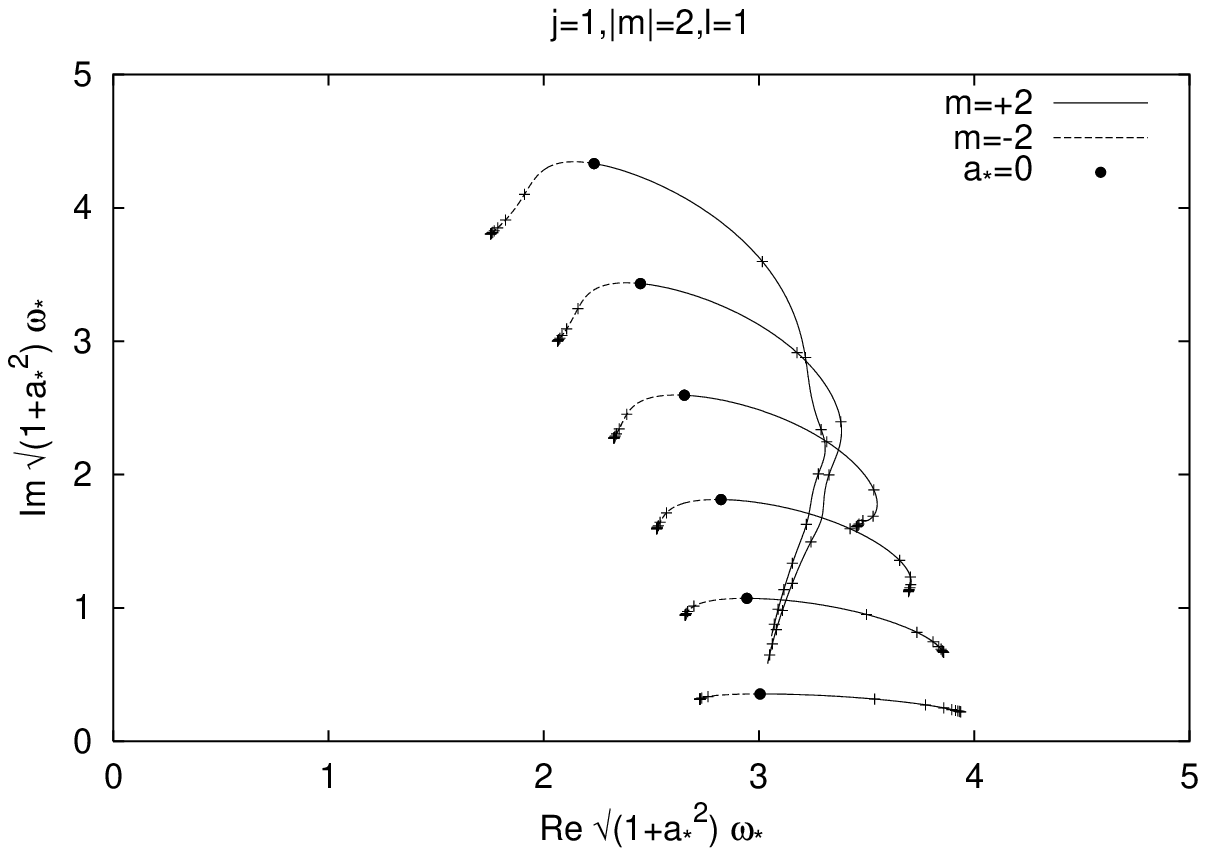}}
\caption{The first six quasi-normal modes of Kerr black hole for $j=1$, $m=\pm2$, $\ell=1$ are plotted in the $\sqrt{1+a_*^2}\omega_*$ plane.
The bullets are the quasi-normal modes of Schwarzschild black hole ($a_*=0$).
The right part starting from the Schwarzschild mode is the branch of the $m=+2$ modes.
The left part is the branch of the $m=-2$ modes.
The crosses are the modes with integer values of $a_*$.
}
\label{fig:121}
\end{figure}

\begin{figure}
\centerline{\epsfbox{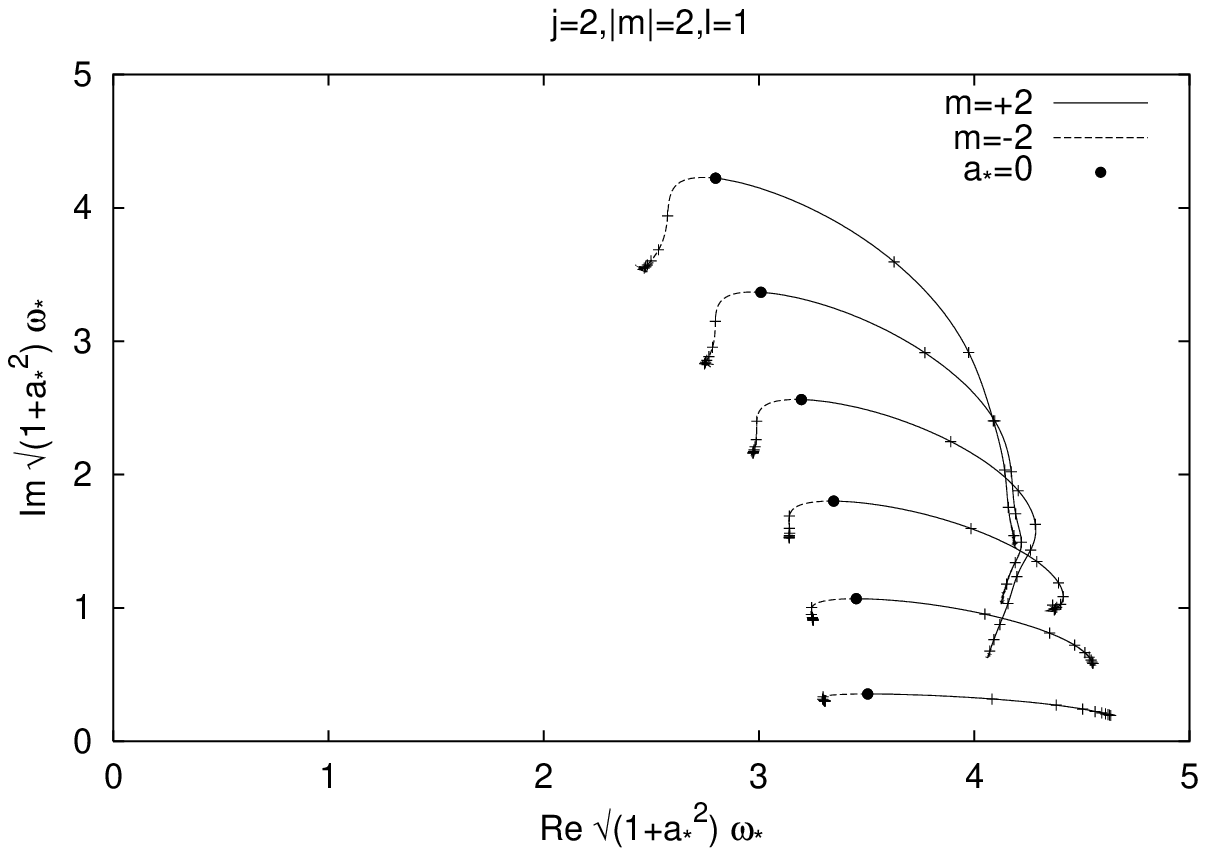}}
\caption{The first six quasi-normal modes of Kerr black hole for $j=2$, $m=\pm2$, $\ell=1$ are plotted in the $\sqrt{1+a_*^2}\omega_*$ plane.
The bullets are the quasi-normal modes of Schwarzschild black hole ($a_*=0$).
The right part starting from the Schwarzschild mode is the branch of the $m=+2$ modes.
The left part is the branch of the $m=-2$ modes.
The crosses are the modes with integer values of $a_*$.
}
\label{fig:221}
\end{figure}

Typical examples of quasi-normal modes of the five-dimensional
Kerr black holes are plotted in Figs~\ref{fig:011}--\ref{fig:221},
where the Kerr parameter $a_*$ is continuously varied.
Here the resonant frequencies are normalized by $\sqrt{\mu}$ rather than $r_H$, 
{\it i.e.}, we work on $\sqrt{1+a_*^2}\omega_*$-plane.
Though the Kerr parameter $a_*$ is taken to be sufficiently large values, each mode
does not seem to pass through the real axis of the frequency plane in any cases.
The $m>0$ branches seem to approach to some points on the real axis, which is
a similar behaviour to the four-dimensional case~\cite{Leaver,Detweiler,Onozawa,Seidel-Iyer}.
Though we have searched for unstable modes up to $|m|,j\le 5$, they are not found.

\section{Summary}
We have investigated the massless scalar field equation in the higher-dimensional Kerr 
black hole background.
The field equation is separable in arbitrary dimensions.
Under the quasi-normal boundary condition, we have searched for the resonant frequencies
in the complex $\omega_*$-plane in the case of the five dimensions.

In the Schwarzschild case ($a_*=0$), the resonant frequencies are
characterized by $\ell$ and the excitation number $N$.
For nonzero Kerr parameter, these resonant frequencies split for
different sets of $m$ and $j$.
Continuously varying the Kerr parameter $a_*$, 
we keep track of resonant modes,
of which trajectories are characterized by the set of wave numbers:
($\ell,m,j,N$).
Each trajectory does not pass through the real axis of the complex frequency plane.
For $m>0$, however, these trajectories seem to approach
to points on the real axis of the frequency plane.
The corresponding time scale of the resonant oscillation in the limit 
$a_*\to 0$,
in which case the horizon radius shrinks to zero, is characterized by the mass
scale $\sim\mu^{1/2}$ rather than the light-crossing time $\sim r_H$.

These results show the evidence of the stability of
 the scalar perturbation of the five-dimensional 
Kerr black holes.

\section*{Acknowledgments}
The authors would like to thank
R.~Emparan,
H.~Ishihara,
H.~Kodama,
T.~Mishima,
K.~Nakamura,
K.~Nakao,
K.~Oda,
H.~Onozawa,
M.~Sasaki
and
T.~Shiromizu 
for useful discussions and comments.
D.I. was supported by JSPS Research, and this research was
supported in part by the Grant-in-Aid for Scientific Research Fund (No. 6499).



\begin{thebibliography}{99}
\bibitem{brane}
N. Arkani-Hamed, S. Dimopoulos and G.R. Dvali,
Phys. Lett. {\bf B429}, 263 (1998);
I. Antoniadis {\em et al.}, Phys. Lett. {\bf B436}, 257 (1998);
L. Randall and R. Sundrum, Phys. Rev. Lett. {\bf 83}, 3370 (1999).

\bibitem{LHC1}
P.C. Argyres, S. Dimopoulos and J. March-Russell, 
Phys. Lett. {\bf B441}, 96 (1998).

\bibitem{LHC2}
S.B. Giddings and S. Thomas, hep-ph/0106219.

\bibitem{Myers-Perry}
R.C. Myers and M.J. Perry, Ann. Phys. {\bf 172}, 304 (1986).  

\bibitem{unique}
G.W. Gibbons, D. Ida and T. Shiromizu, gr-qc/0203004;
G.W. Gibbons, D. Ida and T. Shiromizu, Phys. Rev. Lett. {\bf 89}, 041101 (2002);
G.W. Gibbons, D. Ida and T. Shiromizu, Phys. Rev. D {\bf 66}, 044010 (2002);
M. Rogatko, Class. Quantum Grav. {\bf 19}, L151 (2002).

\bibitem{ring}
R. Emparan and H.S. Reall, Phys. Rev. Lett. {\bf 88}, 101101 (2002).

\bibitem{topology}
M. Cai and G.J. Galloway, Class. Quantum Grav. {\bf 18}, 2707 (2001).

\bibitem{frolov}
V.~Frolov and D.~Stojkovi\'{c} , gr-qc/0211055


\bibitem{Leaver}
E.W.~Leaver , Proc. R. Soc. London. {\bf A402}, 285 (1985)

\bibitem{Leaver2}
E.W.~Leaver , J. Math. Phys. {\bf 27} 1238 (1986)

\bibitem{Detweiler}
S.L.~Detweiler and S.~Chandrasekhar , Proc. R. Soc. London. {\bf A352}, 381 (1977)

\bibitem{Detweiler-Ipser}
S.L.~Detweiler and J.~R.~Ipser, Astrophys.\ J.\  {\bf 185}, 675 (1973).

\bibitem{Onozawa}
H.~Onozawa , Phys. Rev. D {\bf 55}, 3593 (1997)

\bibitem{Gautschi}
W.~Gautschi , SIAM Rev. {\bf 9}, 24 (1967)

\bibitem{Press-Teukolsky}
W.H.~Press and S.~A.~Teukolsky , Astrophys.\ J.\  {\bf 185}, 649 (1973)

\bibitem{Seidel-Iyer}
E.~Seidel and S.~Iyer , Phys. Rev. D {\bf 41}, 374 (1990)

\end{thebibliography}
\end{document}